# Broadband planar electromagnetic hyper-lens with uniform magnification in air

Ran Sun[1, #], Fei Sun[1, #, *], Hanchuan Chen[1], Yichao Liu[1] and Qi Wang[1]

*1 Key Lab of Advanced Transducers and Intelligent Control System, Ministry of Education and Shanxi Province, College of Electronic Information and Optical Engineering, Taiyuan University of Technology, Taiyuan, 030024 China*

[#] Ran Sun and Fei Sun contributed equally to this work.

[*] Corresponding author: sunfei@tyut.edu.cn

**Abstract**

A planar hyper-lens, capable of creating sub-wavelength imaging for broadband electromagnetic wave, is designed based on electromagnetic null medium. Subsequently, a scheme for the implementation of the proposed hyper-lens is given by using well-designed flexural metal plates, which function as the reduced electromagnetic null medium for TM-polarized microwaves. Both simulated and measured results verify that the hyper-lens designed with flexural metal plates can achieve super-resolution imaging for microwave at operating wavelength ($\lambda_0$=3cm) with a resolution of $0.25\lambda_0$ and a uniform magnification of about 5. Moreover, the designed hyper-lens ensures that both the object and image surfaces are planes, and simultaneously provides a uniform magnification for objects in different positions. Additionally, the proposed hyper-lens offers a broadband super-resolution imaging capabilities, achieving good super-resolution imaging effects for microwave frequencies ranging from 8.5 to 11 GHz. The proposed hyper-lens may find applications in high precision imaging, detection, and sensing.

**Introduction**

The resolution of conventional optical systems is limited by the diffraction limit, which restricts the resolution to features that are larger than about half the wavelength of light [1]. The diffraction limit arises from the loss of evanescent waves, which contain detailed information of the object but cannot reach the far-field image plane due to their exponential decaying [2]. To beat the diffraction limit and achieve super-resolution imaging, various near-field scanning methods have been proposed, such as near-field scanning optical microscopy [3], scanning electron microscope [4], atomic force microscope [5] and fluorescence imaging microscopy [6]. Despite these advancements, encounter challenges remain with these near-field scanning methods, such as the inability to capture real-time dynamic images and the risk of sample damage during the scanning process.

To address these challenges, some special lenses with sub-wavelength resolutions for electromagnetic (EM) waves have been proposed. Noteworthy among these are EM super-lenses with negative refractive indices [7, 8] and EM hyper-lenses with highly anisotropic media [9-11]. Nevertheless, due to the high loss and narrow resonance range of negative refractive index materials, the imaging performance of super-lenses is often worse than that of hyper-lenses. This discrepancy has led to a heightened focus on studies aimed at realizing hyper-lenses for various waves in different frequencies [12-15] and improving the imaging quality of hyper-lenses [16-23]. However, most of the current EM hyper-lenses have non-planar object/image surfaces [9-12, 14, 15], making them difficult to integrate with subsequent planar detection structures. Although many methods have been proposed to achieve EM hyper-lenses that have both planar object and image surfaces, they inevitably result in non-uniform magnification across hyper-lenses, which varies depending on the object's position [16-19, 21, 22]. To address this issue, in our previous study, we utilized waveguide meta-materials to realize an EM hyper-lens with object and image surfaces that are both planar and exhibit uniform magnification [23]. However, our previously designed EM hyper-lens still suffers from a narrow bandwidth (only near the cut-off frequency of the waveguide) and can only operate in a waveguide environment which limits the range of its applications. Therefore, realizing a broadband EM hyper-lens with planar object and image surfaces and uniform magnification in air, not confined to a waveguide, remains a challenge.

To address the common limitation of current EM hyper-lens mentioned above, a broadband planar EM hyper-lens is proposed in this study, which has both planar object and image surfaces and is capable of achieving super-resolution imaging for TM-polarized EM waves with uniform magnification in air. The proposed hyper-lens is based on the EM null medium [24-28], which performs as a perfect endoscope for TM-polarized EM waves and can be realized by various metallic metamaterials [29-36]. In this study, the hyper-lens is designed as a L-shaped EM null medium and realized by L-shaped copper plates in Fig. 1(a), in which the length of each air channel between adjacent copper plates satisfies the Fabry-Pérot condition. Both simulated and measured results show the proposed hyper-lens can create a super-resolution imaging (e.g., $0.25\lambda_0$) with planar object/image surfaces and uniform magnification in air for broadband EM waves (8.5 to 11 GHz).

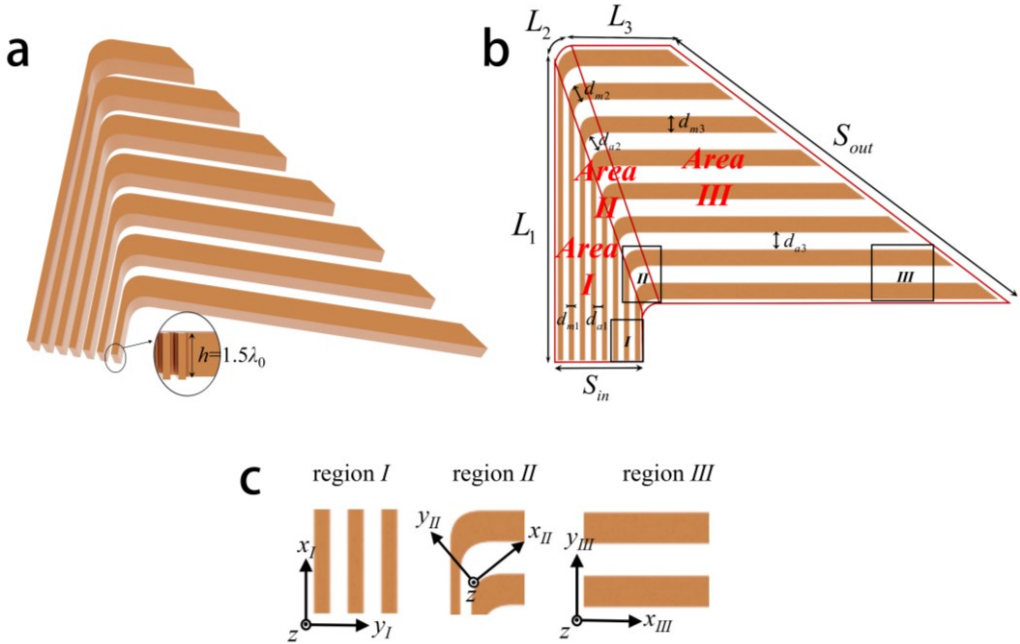

Figure 1. (a) The 3D schematic diagram of the designed hyper-lens. (b) The top view of the designed hyper-lens, which can be divided into three regions I, II, and III due to the different directions of copper plates. (c) The local magnified view of the metal array in (b). The $x$ direction of the local coordinate system is defined as the arrangement direction of the local metal plate. For the three regions I, II, and III in (b), different local coordinate systems can be defined respectively, ensuring that the arrangement direction of the metal plates in each area corresponds to the $x$ direction of the local coordinate system. The schematic of the reduced EM null medium with the main axis along the $x$ direction by metal plates. The subscripts '$m$' and '$a$' denote metal and air, respectively.

**Design Method**

Before designing the planar hyper-lens, we should first review the material parameters and properties of the EM null medium. The ideal EM null medium is a highly anisotropic EM medium obtained through an extremely spatial stretching transformation in transformation optics, where its permittivity and permeability approach infinity along the principal axis, and approach zero in other directions [24, 37]. Taking the ideal EM null medium with the principal axis along the *x*-direction as an example, its permittivity and permeability can be represented as $\varepsilon = \mu = diag\,(1/\Delta,\Delta,\Delta)$, $\Delta \rightarrow 0$. "*diag*" represents a diagonal matrix. The ideal EM null medium can perform as a perfect EM endoscope, which can project the distribution of EM field from one surface to another surface along its principal axis However, the ideal EM null medium is difficult to be realized due to the singularities in the material parameters. Fortunately, the reduced EM null medium can be realized with the help of metallic metamaterials [29-36]. For example, when designed to operate at microwave frequencies, sub-wavelength spaced copper plates can be used to effectively simulate a reduced EM null medium for TM-polarized waves, with the principal axis aligning precisely with the orientation of the copper plates [26, 31].

Unlike the previous EM devices that utilized flat metal plates to create a reduced EM null medium, the hyperlens proposed here is based on flexural metal plates (including the region II corner part in Fig. 1(b)). Next, we will first explain why such flexural metal plates can perform as a reduced EM null medium and why they can achieve a super-resolution imaging effect physically. Then, we will provide a detailed process for designing the hyperlens in Fig. 1.

The hyperlens in Fig. 1(b) can be divided into three regions I, II, and III due to the different directions of the copper plates. Regardless of which region, we can establish a local coordinate system such that the *x*-direction of the local coordinate system aligns with the direction of the metal plates (see Fig. 1(c)). Considering the spacing between the metal plates is on the subwavelength order, i.e., $\lambda_0 \gg (d_a+d_m)$, where $\lambda_0$ is the working wavelength, $d_a$ and $d_m$ are the thickness of the air and metal, respectively, the array of metal plates in each local coordinate system can be treated as an equivalent medium, whose effective relative permittivity and permeability can be obtained based on the effective medium theory [38-39]: $\varepsilon_x \rightarrow \infty$, $\varepsilon_y = 1/f_a$, $\mu_z = f_a$, where $f_a=d_a/(d_a+d_m)$ is filling factor of the air (here only show the EM parameters that are effective for TM-polarized waves). Note that the metal is modeled as perfect electric conductor (PEC) when operating at microwave frequencies. If the filling factor is designed as $f_a=0.5$, the equivalent relative permittivity and permeability of the array of metal plates in each local coordinate system can be further expressed as: $\varepsilon_x \rightarrow \infty$, $\varepsilon_y = 2$, $\mu_z = 0.5$. In this case, the permittivity is infinitely large in the local *x*-direction and relatively small in the local *y*-direction, the magnetic permeability is smaller than 1 in the local *z*-direction, which can be treated as a reduced EM null medium with the main axis along the local *x*-direction for TM-polarized EM wave. The theoretical analysis indicates that for the flexural metal plates in Fig. 1(b), each local region can be considered as a reduced EM null medium with its principal axis aligned with the orientation of the metal plates. Furthermore, due to the continuous gradient in the orientation of the flexural metal plates, they can be treated as an equivalent reduced EM null medium with a continuously varying principal axis.

For TM-polarized EM waves, the magnetic field is oriented in the *z*-direction. At the same time, for EM waves in the microwave band, the metal behaves as a PEC with a skin depth that is essentially zero, so within each subwavelength air channel between two metal plates, the boundary conditions require the electric field to always be perpendicular to the channel (i.e., the electric field is always oriented in the *y*-direction of the local coordinate system). In such a case, the EM wave propagation within the metal plate array always maintains the fundamental transverse electromagnetic (TEM) mode. Therefore, metal plates sample incoming EM wave, and transmit them to the TEM mode in each channel. In this scenario, if the length of each metal plate channel satisfies the Fabry-Pérot resonance condition (that is, $L=m\lambda_0$, $m = 1,2,3...$), the transmission coefficient becomes ±1 (indicating the absence of a reflected wave) for incoming EM wave with any incidence angle, including evanescent waves [40]. From a physical perspective, the subwavelength metal plate arrays' ability to sample and transmit EM waves through discrete channels for subsequent synthesis is analogous to the canalization phenomenon observed in EM waves [40-42]. This phenomenon arises from the extreme anisotropy of the reduced EM null medium and is also the reason behind the achievement of subwavelength resolution imaging.

Unlike previous studies which featured subwavelength metal plates in straight configurations without turns [24-26, 40-42], the hyperlens in this study introduces a continuous variation in the orientation of the metal plate channels. Additionally, as the metal plate array incorporates turns, there are also minor adjustments to the geometric dimensions of each metal plate that forms the channel. In accordance with the boundary conditions, the EM mode in each subwavelength waveguide consistently operates in the fundamental TEM mode. Consequently, even with minor variations in the orientation and dimensions of each subwavelength metal channel, the mode within each channel remains the fundamental TEM mode. This characteristic ensures that the canalization phenomenon can always be produced, thereby enabling the hyperlens depicted in Fig. 1(b) to achieve a super-resolution imaging effect.

Following the above analysis of the super-resolution imaging physical mechanism, the design of the hyperlens can be transformed into the issue of how to connect the plane object and image surfaces of different geometric sizes with channels formed by metal plates, and to ensure that the length of each channel satisfies the Fabry-Pérot condition. Here, we outline the following design process to achieve the hyper-lens in Fig. 1 First, to achieve planar hyper-lens with magnification, the input surface and the output surface should be designed as two planes, and the width of the output plane $S_{out}$ should be larger than the width of the input plane $S_{in}$, i.e., $S_{out}>S_{in}$. Subsequently, to connect two planes with different widths $S_{out}$ and $S_{in}$ by metal plates, the metal plates can be designed into three regions in Fig. 1(b). For simplicity, the regions I and III use flat metal plates directly connecting the input and output surfaces, respectively. Region II is a rounded corner with a radius of $d_{m3}$ and an arc length of $L_2=\pi d_{m3}/2$, smoothly connecting the subwavelength air channels in other two regions to reduce reflections. Next, to ensure that each air channel can be connected from the input surface to the output surface and obtain the best transmittance, the number of metal plates $N$ (also the number of air channels $N$-1) across the three regions is kept the same. For the whole structure to behave as a reduced EM null medium, the conditions that hold for effective medium theory, i.e., $(d_m+d_a) \leq \lambda_0/4$ should be satisfied. Considering the geometric relations $S_{in}=(Nd_{m1}+(N-1)d_{a1})$, $S_{out}=(Nd_{m3}+(N-1)d_{a3})$, the total number of metal plates $N$ is indirectly constrained by the requirement: $N \geq \max\{4(S_{in}+d_{a1})/\lambda_0, 4(S_{out}+d_{a3})/\lambda_0\}$. Then, to make the equivalent medium in each region locally behave as the reduced EM null medium the filling factors in three regions are chosen as $f_m=f_a=0.5$ (i.e., $d_{m1}=d_{a1}$, $d_{m2}=d_{a2}$, and $d_{m3}=d_{a3}$). Furthermore, there are some geometrical relationships between the input/output surfaces of the lens and the widths of the air channels: $(S_{in}+d_{m1})/N=2d_{m1}$, $(S_{out}+d_{m3})/N=2d_{m3}$. The air channel width in region I $d_{a1}=d_{m1}$ determines the resolution, while the air channel width in region III $d_{a3}=d_{m3}$ determines the magnification. Finally, to ensure that the whole structure has no reflection (maximizing transmittance), the total length of each metal plate and air channel needs to satisfy the Fabry-Pérot resonance condition $L_1+L_2+L_3=m\lambda_0$, where $m$ is an integer (1, 2, 3, ...).

**Simulated Results**

To verify performance of the hyper-lens for EM, numerical simulations are conducted by COMSOL Multiphysics 5.6 with the license number 9406999. All simulations are 2D cases where the wave optics module with steady-state solver are selected to simulate the EM field distribution, and the free tetrahedral meshings with the maximum grid $\lambda_0/10$ are used. The working wavelength is designed as $\lambda_0=3$cm.

As an example, the parameters of a hyper-lens are designed as $S_{in}=13\lambda_0/8$, $S_{out}=65\lambda_0/8$, $N=7$, $h=1.5\lambda_0$, $d_{m1}=d_{a1}=\lambda_0/8$, $d_{m3}=d_{a3}=3\lambda_0/8$, $L_2=\pi d_{m3}/2\approx 0.6\lambda_0$, and $L_1+L_2+L_3=8\lambda_0$. When two EM sources with the subwavelength separation $d_{in}=0.25\lambda_0$ are located at the input surface $S_{in}$ of the hyper-lens, the normalized amplitude of the simulated magnetic field's $z$ component is shown in Figs. 2(a). In this case, the normalized amplitude of magnetic field s $z$ component intensity($|H_z|$) on the input surface $S_{in}$ and the output surface $S_{out}$ of the designed hyper-lens are also plotted in Figs. 2(b), which further show two sub-wavelength spaced sources $d_{in}=0.25\lambda_0$ on the input surface can be magnified to obtain two distinguishable images with a magnified spacing $d_{out}=0.131\lambda_0$ on the output surface, where the blue curves are lines of the output surface $S_{out}$ and the red curves are lines of the input surface $S_{in}$. In this example, the magnification factor $M=d_{out}/d_{in}=5.24$ is observed. Next, we will show the magnification remains the same when the positions of two sources change in the input surface, i.e., uniform magnification factor.

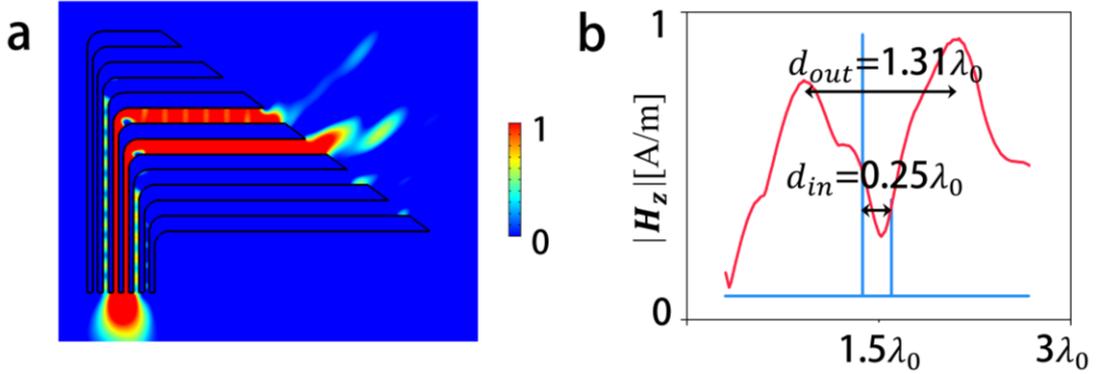

Figure 2. (a) The normalized amplitude distribution of the simulated magnetic field's $z$-component $|H_z|$, when two EM sources with a subwavelength separation of $d_{in}=0.25\lambda_0$ are on the input surface $S_{in}$ of the designed hyper-lens. (b) The red and blue curves are the normalized magnetic field's $z$ component $|H_z|$ at the output surface $S_{out}$ and the input surface $S_{in}$, respectively. In this example, the $\lambda_0=3$cm, $d_{m1}=d_{a1}=\lambda_0/8$, $d_{m3}=d_{a3}=3\lambda_0/8$, $L_2=\pi d_{m3}/2\approx 0.6\lambda_0$, $L_1+L_2+L_3=8\lambda_0$.

To verify the uniform magnification of the designed hyper-lens, additional simulations are given in Fig. 3. Figs. 3(a)-(e) illustrate that the normalized amplitude of magnetic field's $z$-component as the locations of two-line current sources vary on the input surface of the designed hyper-lens. During the location-scanning of the object points, the distance between the two EM sources of is kept constant at $0.25\lambda_0$, and each scan translates the two EM sources together by $0.25\lambda_0$ along the horizontal direction. Note that each scan requires two sources to be located at the entrances of air channels (rather than the metal channels) to ensure EM information can be transmitted to the output surface $S_{out}$ through the air channels. The corresponding magnification factors can be calculated from the simulated results shown in Figs. 3(a)-(e), which are presented in Fig. 3(f). The simulation results demonstrate that the designed hyper-lens not only has planar input and output surface, but also exhibits a magnification factor that remains stable at approximately 5.15, regardless of the source's position on the input plane, indicating that uniform EM magnification has been achieved.

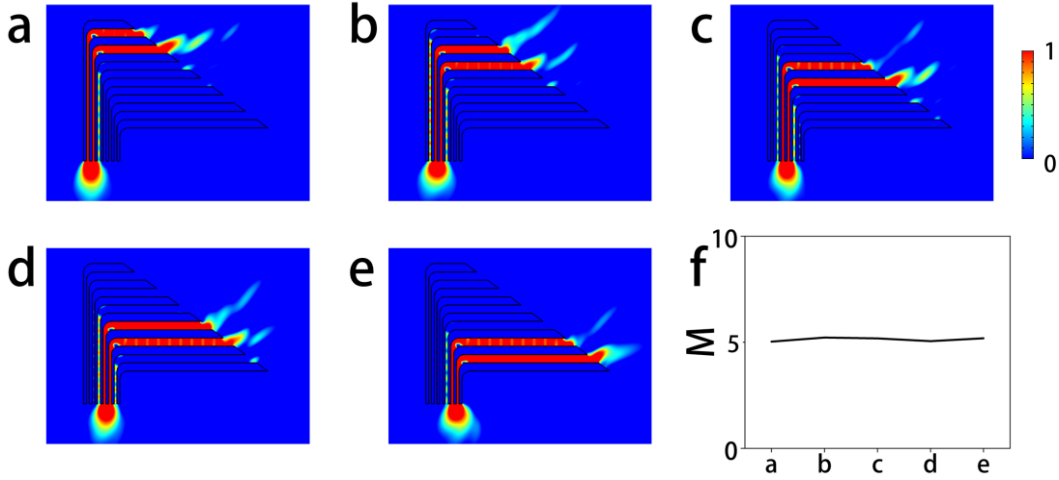

Figure 3. (a)-(e) The normalized amplitude distribution of the simulated magnetic field's z-component as the locations of two EM sources on the input surface $S_{in}$ of the designed hyper-lens change. The separation between the two EM sources is kept constant at $d_{in}=0.25\lambda_0$ during the location-scanning process. (f) The corresponding magnification factor $M=d_{out}/d_{in}$ for the cases in Figs. 3(a)-(e), with the horizontal coordinates $x$ representing the image serial number. The parameters of the hyper-lens and other numerical settings are consistent with those in Fig. 2.

From a physical perspective, although many previous studies have shown that subwavelength metal plates can achieve complete transmission (no reflection) for EM waves at any incident angle, including evanescent waves, only when the wavelength meets the Fabry-Pérot resonance condition, this does not preclude the achievement of super-resolution effects from an engineering application standpoint. Even if the incident wavelength deviates from the Fabry-Pérot resonance, causing imperfect transmission at some angles and for some evanescent wave components, it does not negate the potential for super-resolution. Indeed, recent studies have demonstrated that such subwavelength metal plates can achieve broadband cloaking [31] and concentrating [36]. Consequently, we numerically study the imaging performance of the designed hyper-lens in Fig. 1 when the wavelength of EM sources $\lambda$ deviates from the designed wavelength $\lambda_0=3$cm, which corresponds to the operating frequencies of the EM wave $f_0 = 10.0$ GHz. The simulated results presented in Figs. 4(a)-(f) demonstrate that two EM sources with sub-wavelength separation $d_{in}=0.25\lambda_0$ can be resolved and magnified at the output surface when the frequencies of EMwaves vary within the range from 8.5GHz to 11.0GHz. The numerical simulations depicted in Fig. 4 illustrate that the designed hyper-lens can still function as an EM planar hyperlens with uniform magnification, delivering a high-quality super-resolution imaging effect over a broadband frequency range centered around the designed wavelength.

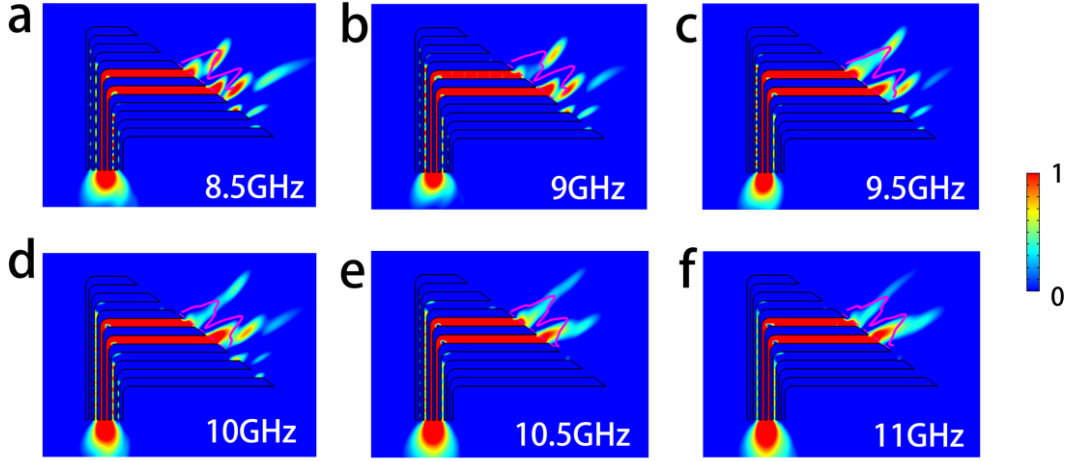

Figure 4. (a)-(f) The normalized amplitude distribution of the simulated magnetic field's $z$-component $|H_z|$, when two EM sources with a subwavelength separation of $d_{in}=0.25\lambda_0$ are positioned on the input surface $S_{in}$ of the designed hyper-lens. The frequencies of EM sources range from 8.5GHz to 11.0GHz. The pink curves represent the normalized $|H_z|$ (magnetic field) distribution. The parameters of the hyper-lens and other numerical settings are consistent with those in Fig. 2.

**Experimental measurements**

Next, a hyperlens sample is fabricated and an experiment is designed to demonstrate the super-resolution imaging capabilities of the proposed hyper-lens. Figs 5(a) show the experimental design diagram for EM waves. The working wavelength is designed as $\lambda_0$=3cm, which corresponds to the working frequency 10GHz. In this experiment, the geometrical parameters of the fabricated sample are chosen as follows: $h=1.5\lambda_0$, $d_{m1}=d_{a1}=\lambda_0/8$, $d_{m3}=d_{a3}=3\lambda_0/8$, $L_2=\pi d_{m3}/2\approx 0.6\lambda_0$, $L_1+L_2+L_3=8\lambda_0$.

For the EM experiment, the hyper-lens sample is placed on a platform made of foam sheet, approximately 20cm above the ground. The EM signal is generated by a VNA (ROHDE&SCHWARZ ZVL13) at its output port 2. It is then connected via a coaxial cable to a power splitter (SQY-PS2-2/18-SE), which divides it into two channels of EM sources, which are further connected to two source loop antennas, respectively, serving as two EM sources to be imaged on the object plane of the hyper-lens. The two EM sources are positioned in the middle of the third and fourth air channels of the object plane, immediately adjacent to it, with a separation distance of $d_{in} = 0.25\lambda_0$. A detection loop antenna is positioned at a distance of $l$= 5mm from the image plane of the hyper-lens sample, aligned with the center of the hyper-lens (i.e., $h/2$) in the $z$-direction. It is controlled by stepper motor and movable along the parallel direction of the image plane at 2mm intervals and is used to measure the distribution of the magnetic field's $z$-component at the image plane of the hyper-lens sample. Subsequently, the detection loop antenna is connected via coaxial cable to port 1 of the VNA for measuring the $S_{21}$ parameters. The entire system is situated within an EM dark room to eliminate external EM signal interference. The obtained $S_{21}$ parameters are subsequently utilized to compute the normalized magnetic field distributions, depicted by the black curves in Fig. 2c. The measured magnetic field (black curve) in Fig. 2c closely matches the simulated result (red curve) in Fig. 2c, which verifies that the fabricated hyper-lens can achieve a subwavelength resolution of $0.25\lambda_0$.

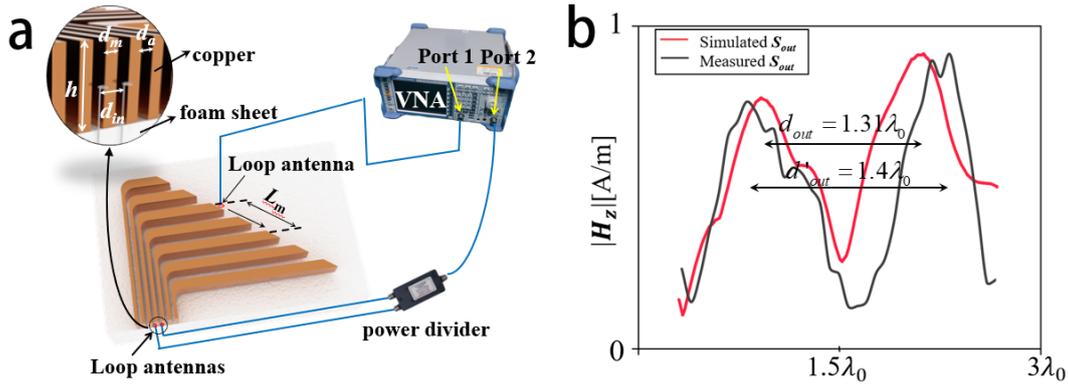

Fig. 5. (a) The experimental design diagrams for EM wave. (b) The measured (black curves) and simulated (red curves) normalized magnetic field distribution at the image plane (indicated by the yellow dashed line in Fig. 2(a)). The $d_{out}$ and $d'_{out}$ labelled in the figure are the peak spacing for simulation and experiment, respectively

## Conclusions

Unlike traditional hyper-lenses, which mainly utilize flat subwavelength spaced metal plate arrays, ensuring that both the object and image planes are planar while maintaining uniform magnification at all positions is challenging. The hyper-lens proposed in this study uses flexural metal plates with smooth bending angles, equivalent to a reduced EM null medium with gradually varying principal axes. Even with turns and slight size variations, it can still maintain the operation of the fundamental TEM mode within each channel formed by the subwavelength metal plates. When all channel lengths meet the Fabry-Perot condition, it can achieve the effect of super-resolution imaging with uniform magnification. Both numerical simulations and experimental measurements demonstrate that the proposed hyper-lens can achieve a resolution of $0.25\lambda_0$ and a uniform magnification of about 5. Furthermore, the hyper-lens exhibits broadband super-resolution imaging in the microwave band from 8.5 to 11 GHz. This work addresses the main limitations of current hyper-lenses. Firstly, the designed hyper-lens can have planar object and image planes while maintaining uniform magnification at any position. Secondly, the designed hyper-lens does not require inhomogeneous medium and can operate in air without the need for a specific environment such as a waveguide. Thirdly, the designed hyper-lens can achieve broadband super-resolution imaging effects.

**Competing interests**
The authors declare no competing financial interests.

**Data availability**
The main data and models supporting the findings of this study are available within the paper. Further information is available from the corresponding authors upon reasonable request.

**Acknowledgments**
This work is supported by the National Natural Science Foundation of China (Nos. 12274317, 12374277, and 61971300), Basic Research Project of Shanxi Province 202303021211054, and University Outstanding Youth Foundation of Taiyuan University of Technology.